\documentclass[aps,prd,showpacs,amsmath,amssymb]{revtex4}

\usepackage{slashed}
\usepackage[dvips]{color}
\usepackage[dvips]{epsfig}
\usepackage{latexsym}
\usepackage{bm}
\usepackage{upgreek}
\usepackage{mathrsfs}
\usepackage{times}
\usepackage{amsthm}
\usepackage{amssymb}
\usepackage{epsfig}
\usepackage{graphicx}
\usepackage{amsmath}

\theoremstyle{plain}

\newtheorem{lemma}{Lemma}[section]

\theoremstyle{definition}

\theoremstyle{remark}
\newtheorem{remark}[lemma]{Remark}

\include{graphics}
\begin{document}

\title{The Nambu sum rule and the relation between the masses of composite Higgs bosons}

 \author{G.E.~Volovik}
\affiliation{Low Temperature Laboratory, School of Science and Technology, Aalto University, Finland}
\affiliation{L.D. Landau Institute for Theoretical Physics, Moscow, Russia}

\author{M.A.~Zubkov}
\affiliation{ITEP, B.Cheremushkinskaya 25, Moscow, 117259, Russia }

\date{\today}
\begin{abstract}
We review the known results on the bosonic spectrum in various NJL models both in the condensed matter physics and in relativistic quantum field theory including $^3$He-B, $^3$He-A, the thin films of superfluid He-3, and QCD (Hadronic phase and the Color Flavor Locking phase). Next, we  calculate bosonic spectrum in the relativistic model of top quark condensation suggested in \cite{Miransky}.
  In all considered cases the sum rule appears that relates the masses (energy gaps) $M_{boson}$ of the bosonic excitations in each channel with the mass (energy gap) of the condensed fermion $M_f$ as $\sum M_{boson}^2 = 4 M_f^2$. Previously this relation was established by Nambu in \cite{Nambu} for $^3$He-B and for the s - wave superconductor. We generalize this relation to the wider class of models and call it the Nambu sum rule. We discuss the possibility to apply this sum rule to various models of top quark condensation. In some cases this rule allows to calculate the masses of extra Higgs bosons that are the Nambu partners of the $125$ GeV Higgs.
\end{abstract}



\maketitle


\section{Introduction}

It is difficult to overestimate the role of the NJL approximation in field theory (i.e. the approximation with the effective 4 - fermion interaction) \cite{NJL}. It gives qualitative understanding of the  formation of fermion condensates in a number of models that describe various physical problems from superconductivity and superfluidity \cite{Nambu} to top quark condensation \cite{topcolor1}. However, any NJL model is only a low energy approximation to the  microscopic theory.
The NJL models are not renormalizable. Therefore, they are to be considered as the phenomenological models with the finite ultraviolet cutoff $\Lambda$. In most of the papers on the NJL models the physical quantities are evaluated in one - loop approximation (i.e. in the leading order in $1/N$ expansion). It is worth mentioning, that formally the contributions of higher loops to various physical quantities may be strong. For example, in \cite{cveticreview,cvetic} it has been shown that the next to leading (NTL) order approximation to the  fermion mass $M_T$ in the simplest model of top quark condensation is weak compared to the one - loop approximation only if $M_T \sim \Lambda$. Actually, all dimensional parameters of the relativistic NJL models (calculated nonperturbatively or taking into account higher orders of the perturbation theory) are typically of the order of the cutoff unless their small values  are protected by symmetry.

Nevertheless, there is another way to look at the NJL models. We can consider the one - loop approximation for the calculation of various quantities like fermion and boson masses (i.e. the mean field approximation, or the leading order in $1/N$ expansion). The higher loops are simply disregarded. This is usually done in the NJL approximation to QCD \cite{NJLQCD} or Technicolor \cite{Simmons}, where all dimensional parameters are of the order of the ultraviolet cutoff so that the corrections to the leading order $1/N$ approximation are not so large. (For the results of the the nonperturbative numerical lattice investigation of the NJL model see, for example, \cite{latticeNJL}).  However, this is also done in many papers on the models of top quark condensation (TC) \cite{Miransky,topcolor1}, where the cutoff is assumed to be many orders of magnitude larger than the mass of the top quark. There were only a few papers on the next to leading order approximation (see, for example, \cite{cvetic,cveticreview}). Besides, in the evaluation of the Standard Model fermion masses in the Technicolor theory due to the Extended Technicolor (ETC) interactions \cite{Simmons} the effect of the ETC is taken into account through the effective four - fermion term. No loop contributions due to this term are considered. However, those loop contributions would give values of masses $\sim \Lambda^2_{ETC}, \Lambda_{ETC} \gg M_T$. The justification may be based on the assumption that we deal with the phenomenologial model that is to be considered at the tree level (ETC), or in one loop (TC) without taking into account higher loop contributions. However, the more rigorous explanation is that there exist the contributions of the microscopic theory  due to the trans - $\Lambda$ degrees of freedom that are not taken into account in the NJL approximation. Those contributions cancel the dominant higher loop divergences. Therefore, the one - loop results (TC) and tree - level results (ETC) dominate. (See also discussion in section \ref{NJL1loop}.) In this paper we assume that this pattern takes place in the models of top - quark condensation. This means that there exist the contributions to the Higgs boson masses and to the quark masses that come from the energies larger than $\Lambda$ and are not accounted by the NJL model. They are assumed to cancel the quadratic divergent contributions to the (squared) Higgs boson masses and linear divergent contributions to the quark masses. The consideration of the possible  mechanisms that may provide this are out of the scope of the present paper. We only mention that there exist the situation, when such a pattern is realized. Namely, in quantum hydrodynamics there formally exist the divergent contributions to  various physical quantities  (for example, to the vacuum energy). Nevertheless, the hydrodynamics may be considered with these divergent contributions subtracted, and this is how classical hydrodynamics appears as a low energy approximation to quantum theory. The origin of this cancellation is well - known \cite{quantum_hydro}. It is provided by the thermodynamical stability of vacuum. Recently it was suggested that the similar mechanism is responsible for the cancellation of the ultraviolet divergences in vacuum energy (quantum gravity) and for the cancellation of the quadratically divergent contribution to the Higgs boson mass in the Weinberg - Salam model \cite{hydro_gravity}.

Basing on this assumption we expect that in the relativistic models of top - quark condensation quantitative predictions of the one - loop NJL approximation may be as accurate as in the BCS models of superconductivity or superfluidity. It is worth mentioning, that in  microscopic theories of top quark condensation there is no confinement (otherwise the top quark would be confined into the regions of space smaller than $1$ TeV $^{-1}$). In this aspect these theories differ essentially from  QCD, where the absence of confinement in NJL approximation does not allow to use it widely for the consideration of low energy physics. (For the attempts to include the description of confinement to NJL approximation see \cite{NJLconf}.)

More specifically, we investigate the particular case of the NJL model suggested in \cite{Miransky}.  We calculate its bosonic spectrum and establish the relation between the masses of bosonic excitations and the fermion masses. This relation is similar to the relation found in $^3$He-B and in the s - wave  superconductors between the energy gaps of the scalar excitations and the fermion energy gap. This relation was first noticed in \cite{Nambu} by Nambu.  In the form
 $M_{1}^2 + M_{2}^2 = 4 \Delta^2$ it is valid in the effective NJL - like model  of $^3$He-B for the boson energy gaps $M_{1,2}$ existing at each value of  $J = 0,1,2$, where  $J$ is the quantum number corresponding to the total angular momentum of the Cooper pair. It relates them to the constituent mass of the fermion excitation $\Delta$ existing due to the condensation.
 The similar relation was also discussed in the Nambu - Jona - Lasinio approximation \cite{NJL} of QCD, where  it relates the $\sigma$ - meson mass and the constituent quark mass $M_{\sigma} \approx 2 M_{quark}$.  (In the non-relativistic BCS theory the role of the masses of the fermionic and bosonic excitations is played by the energy gaps in the fermionic and bosonic spectrum respectively.) Recent discussion of Higgs modes in condensed matter systems can be found in
Refs. \cite{Endres2012,Barlas2012,Gazit2012,KunChen2013,Tsuchiya2013}
and in references therein.

 We introduce the notion of the Nambu sum rule that is the generalization of the mentioned above relations to the theories with condensed fermions (such that there is the single fermion, whose constituent mass $M_f$ is essentially larger than the masses of the other fermions). This sum rule reads:
 \begin{equation}
 \sum M_{H, i}^2  \approx 4 M^2_f, \label{NSR}
\end{equation}
In the l.h.s. of this equation the sum is within the given channel over the composite scalar excitations such that the mentioned fermion with mass $M_f$ contributes to their formation.  We do not give here the general proof of this sum rule. Instead, we consider several models, where it holds.
In addition to our results on the bosonic spectrum of the mentioned above top quark condensation model we review several NJL models, where the bosonic spectrum is already known.



Recall that the recent experimental results \cite{CMSHiggs, ATLASHiggs} on the $125$ GeV
Higgs exclude the appearance of the other Higgs bosons within the wide ranges
of masses (approximately from $130$ GeV to $550$ GeV). However, this announced
exclusion is related only to the particle with the same cross - section as the only
standard Higgs boson of the Standard Model. The particles that have the smaller
cross - sections are not excluded. To be more explicit, we refer to the recent
data of CMS collaboration \cite{CMSexc}. On Figure 4 the solid black curve
separates the region, where the scalar particles are excluded (above the curve)
from the region, where they are not excluded. For example, the particle with
mass around $200$ GeV and with the cross section about $1/3$ of the Standard
Model cross section is not excluded by these data. The similar exclusion curve
was announced by ATLAS (plenary talk \cite{ATLASexc} at ICHEP 2012, slide 34).

This is the analogy with the superconductivity and superfluidity that prompts that the Higgs
boson may be composite.  (See \cite{Englert,Higgs,Kibble} for the foundation of the Higgs mechanism
in quantum field theory.) In our opinion
the models of the top quark condensation \cite{topcolor1,topcolor,Marciano:1989xd,cveticreview,Miransky,modern_topcolor,Simmons,Hill:1996te,Buchalla:1995dp} are of especial
interest as they relate the Higgs boson to the only known Fermi particle
(top - quark) with the mass of the same order as the Higgs boson mass. Therefore, we have in mind the pattern of top quark condensation dealing with the Nambu sum rule, and in the r.h.s. of Eq. (\ref{NSR}) the top quark mass stands. For the review of the conventional technicolor we refer to  \cite{Weinberg:1979bn,Susskind:1978ms,Simmons,Chivukula:1990bc}. The so - called topcolor assisted technicolor that combines both technicolor and topcolor ingredients was considered, for example, in \cite{Hill:1994hp, LaneEichten, Popovic:1998vb, Braam:2007pm,Chivukula:2011ag}.
For the related models based on the extended color sector see \cite{top_coloron,top_coloron1}  and references therein. The top - seesaw mechanism was considered in \cite{topseesaw}. We also mention the attempts to consider the recently found $125$ GeV resonance as a top - pion \cite{modern_topcolor}.

An interesting consequence of the Nambu sum rule with the top  - quark mass is that if there are only two states in the given channel, then the partner of the $125$ GeV Higgs should have the mass around $325$ GeV.
 It is worth mentioning that in 2011 the
CDF collaboration \cite{CDF} has announced the preliminary results on the excess of events in $Z Z \rightarrow l l \bar{l} \bar{l}$ channel at the invariant mass $\approx 325$ GeV.  CMS collaboration also reported a small excess in this region \cite{CMS}. Although the particle with the cross sections of the Standard Model Higgs is excluded at this mass, this exclusion does not work for the particles with smaller cross sections.  Originally the mentioned excess of events was treated as a statistical fluctuation. However,  in \cite{325,320} it was argued that it may point out to the possible existence of  a new scalar particle with mass $M_{H2}\approx 325$ GeV.

The paper is organized as follows.
 In Section II we consider the condensed matter NJL models of He-3. In subsection II.A we review the hydrodynamic action for He-3. Next, in subsection II.B we consider bosonic spectrum in the NJL model of $^3$He-B.   (This model was considered originally in this respect by Nambu). We present the simple method for the calculation of the bosonic spectrum in this model. In principle, this method with some modifications can be applied to the other models of this section, although we do not present the corresponding calculations.
 In subsection II.C we consider the 3D A - phase of the superfluid $^3$He. In this case the fermions are gapless. However, the Nambu sum rule Eq. (\ref{NSR}) works if in its r.h.s. the average of the angle dependent energy gap is substituted.
In subsection II.D we consider 2D thin films of He-3. There are two main phases (a and b), where the Nambu sum rule works within the effective $2D$ four - fermion model similar to that of $^3$He-B.

Section III is devoted to the relativistic NJL models. In subsection III.A we describe the top quark condensation model of \cite{Miransky} and its particular case considered in this paper. In subsection III.B we calculate the bosonic spectrum of the model. In subsection III.C we present the other example of the relativistic model, where the Nambu sum rule holds, i.e. the NJL model of the color superconductor in the so - called Color - Flavor  Locking (CFL) phase. In the corresponding four - fermion effective model there are two different fermionic energy gaps. Both of them are related to the bosonic masses by the relation similar to the ordinary relation between the constituent quark mass and the mass of the sigma - meson. In subsection III.D we consider the analogy with the Veltman identity for the vanishing of the quadratic loop divergencies in the scalar boson masses.


The model of \cite{Miransky} considered in subsections III.A, III.B. suffers from various problems, and a lot of physics is to be added in order to make it realistic. However, this is the first example, when the Nambu sum rule in the nontrivial form appears in the  relativistic model. There may appear the other nontrivial (and more realistic) models of the top - quark condensation (and the other technicolor - like models), where there are several composite Higgs bosons, whose masses are related by the  Nambu sum rule.
We suggest to look for such schemes basing on the analogy with superfluid $^3$He (we refer to the book \cite{VollhardtWolfle1990}
and to the references therein).

\section{Nambu sum rules in Helium - 3 superfluid}
\subsection{"Hydrodynamic action" in $^3$He}

According to \cite{He3} Helium - 3 may be described by the effective theory with the action
\begin{widetext}
\begin{equation}
S = \int dt d^3x \bar{\chi}_s \{i\partial_t + \mu + \frac{1}{2m}\Delta\}\chi_s - \frac{1}{2}\int dt \int d^3x \int d^3y u(x-y) \sum_{s,s^{\prime}} \bar{\chi}_s(x,t) {\chi}_s(x,t) \bar{\chi}_{s^{\prime}}(y,t) {\chi}_{s^{\prime}}(y,t)
\end{equation}
\end{widetext}

Here $\chi$ is anticommuting spinor variable, $s = \pm$, $\mu$ is the chemical potential, $u(x)$ is the interatomic potential. Then, the integration over the "fast" Fermi - fields (i.e., those with the sufficiently large values of momenta) gives the effective action for the modes living near to the Fermi surface.
Assuming imaginary time and the spin-triplet $p$-wave pairing
(i.e. the Cooper pairing in the state with orbital angular momentum $L=1$ and spin angular momentum $S=1$), in the first approximation this effective action can be written as
\begin{equation}
S_{low} = \sum_{p, s} \bar{a}_s(p) \epsilon(p) a_s(p) - \frac{g}{\beta V} \sum_{p, i=1,2,3} \bar{J}_i(p) J_i(p),
\end{equation}
where
\begin{eqnarray}
&&p = (\omega, k), \quad \hat{k} = \frac{k}{|k|}, \\&&
\epsilon(p) = i\omega - v_F (|k|-k_F)\nonumber\\
&&J_i(p) = \frac{1}{2}\sum_{p_1+p_2 = p}(\hat{k}_1 - \hat{k}_2)  a_\alpha(p_2) \sigma_i a_\beta(p_1)\epsilon^{\alpha \beta}\nonumber
\end{eqnarray}

Here $a_{\pm}(p)$ is the fermion variable in momentum space, $v_F$ is Fermi velocity, $k_F$ is Fermi momentum, $g$ is the effective coupling constant.
The authors of \cite{He3} proceed with the bosonization using the following trick.  The unity is substituted into the functional integral that is represented as $1 \sim \int D \bar{c} D c\,  {\rm exp}(\frac{1}{g}\sum_{p, i, \alpha} \bar{c}_{i, \alpha}(p) c_{i, \alpha}(p))$, where $c_{i, \alpha}, (i,\alpha = 1,2,3)$ are bosonic variables.  These variables may be considered further as the field of the Cooper pairs, which serves as the analog of the Higgs field in relativistic theories.  Shift of the integrand in $D \bar{c} D c$ removes the $4$ - fermion term. Therefore, the fermionic integral can be taken. As a result we arrive at the "hydrodynamic" action for the Higgs field $c$:

\begin{equation}
S_{eff} = \frac{1}{g} \sum_{p, i, \alpha} \bar{c}_{i, \alpha}(p) c_{i, \alpha}(p) + \frac{1}{2}{\rm log} \, {\rm Det} M(\bar{c},c), \label{hydrodynamic}
\end{equation}
where
\begin{widetext}
\begin{equation}
M(\bar{c},c) = \left(\begin{array}{cc}(i\omega - v_F (|k|-k_F))\delta_{p_1 p_2} & \frac{1}{(\beta V)^{1/2}}[(\hat{k}_1 - \hat{k}_2)c_{\alpha}(p_1+p_2)]\sigma_{\alpha}\\-\frac{1}{(\beta V)^{1/2}}[(\hat{k}_1 - \hat{k}_2)c_{\alpha}(p_1+p_2)]\sigma_{\alpha}& - (i\omega - v_F (|k|-k_F))\delta_{p_1 p_2}\end{array}\right) \label{matrix}
\end{equation}
\end{widetext}

\subsection{Nambu sum rules in $^3$He-B}

In the B - phase of $^3$He the condensate is formed in the state with
$J=0$, where ${\bf J}= {\bf L}+{\bf S}$ is the total angular momentum of Cooper pair \cite{VollhardtWolfle1990}

\begin{equation}
c^{(0)}_{i \alpha}(p) = (\beta V)^{1/2} C \, \delta_{p0}\delta_{i\alpha} \,.
\end{equation}

This corresponds to the symmetry breaking scheme $G \rightarrow  H$  with the symmetry of physical laws $G=SO_L(3) \times SO_S(3) \times U(1)$ and the symmetry of the degenerate vacuum states $H=SO_J(3)$. The parameter $C$ satisfies the gap equation
\begin{equation}
 0=\frac{3}{g}- \frac{4}{\beta V} \sum_p (\omega^2 + v_F^2(|k|-k_F)^2+4 C^2 )^{-1}
\end{equation}

The value $\Delta = 2 C $ is the constituent mass of the fermion excitation.
There are $18$ modes of the fluctuations $\delta c_{i \alpha}=c_{i \alpha}- c^{(0)}_{i \alpha}$ around this condensate. Tensor $\delta c_{i\alpha}$ realizes the reducible representation of the $SO_J(3)$ symmetry group of the vacuum (acting on both spin and orbital indices).  The mentioned modes are classified by the total angular momentum quantum number $J = 0,1,2$.

According to \cite{He3gauss,He3B} the quadratic part of the effective action for the fluctuations around the condensate has the form:
\begin{equation}
S^{(1)}_{eff} =\frac{1}{g} (u,v) [1 - g \Pi]   \left(\begin{array}{c}{u}\\v \end{array}\right),
\end{equation}
where $\delta c_{i\alpha}(p) = u_{p i \alpha} + i v_{p i \alpha }$, and the polarization operator at $k = 0$ is given by
\begin{equation}
 \Pi = \left(\begin{array}{cc}\Pi^{uu} & 0 \\ 0 & \Pi^{vv}\end{array}\right)
\end{equation}

At each value of $J=0,1,2$ the  modes $u$ and $v$ are orthogonal to each other and correspond to different values of the bosonic energy gaps.

At $k = 0$ the polarization operator can be represented as
\begin{equation}
\Pi(\omega) = \int_{0}^{\infty} d z \frac{\rho(z)}{z+\omega^2}, \label{disp}
\end{equation}

where the spectral function $\rho \sim \sum |F_{Q \rightarrow f f}|^2$, and $|F_{Q \rightarrow f f}|^2$ is the probability that the given mode $Q$ (in case of $^3$He-B the quantum number $Q=J$)  decays  into two fermions.

At $J=0$, the $v$ - bosonic mode is gapless that can easily be obtained using the gap equation.
Also this follows from the fact that this is the Goldstone mode, which comes from the broken $U(1)$ symmetry. Next,  for any $J$
we have  ($ \sqrt{t} = \epsilon_+ +  \epsilon_- ; \, k_+ =  -k_-; \epsilon^2_{\pm} -   v_F^2(|k|-k_F)^2 -  \Delta^2 =0$):
\begin{widetext}
\begin{eqnarray}
\rho^{u}(t) & \sim & \theta(t-4 \Delta^2) \, \sqrt{1 - \frac{4 \Delta^2}{t}} \,{\rm Sp}\, G^{-1}(\epsilon_+,k_+) \, O^{(J)}_u\, G^{-1}(-\epsilon_-,k_-)\, O^{(J)}_u\, \nonumber\\ &\sim&   \sqrt{1 - \frac{4 \Delta^2}{t}} \,[ (t/2  -  \Delta^2) - \eta^{(J)} \Delta^2]\theta(t-4 \Delta^2) \,\nonumber\\
\rho^{v}(t) & \sim &  \theta(t-4 \Delta^2) \, \sqrt{1 - \frac{4 \Delta^2}{t}} \,{\rm Sp}\, G^{-1}(\epsilon_+,k_+) \, O^{(J)}_v\, G^{-1}(-\epsilon_-,k_-)\, O^{(J)}_v\, \nonumber\\ &\sim&   \sqrt{1 - \frac{4 \Delta^2}{t}} \,[ (t/2  -  \Delta^2) + \eta^{(J)} \Delta^2]\theta(t-4 \Delta^2)\label{RHO}
\end{eqnarray}
\end{widetext}

Here

\begin{widetext}
\begin{eqnarray}
&& G^{-1}(\epsilon,k)  = \left(\begin{array}{cc} (\epsilon-v_F(|k|-k_F)) & 2 C (\hat{k}\sigma)\\
-2 C (\hat{k}\sigma)&(\epsilon+v_F(|k|-k_F))\end{array} \right), \,
O_{u,v}^{ij}  = \left(\begin{array}{cc} 0  & \hat{k}_{+}^i \sigma^j\\
\mp\hat{k}_{+}^i \sigma^j& 0 \end{array} \right), \nonumber\\ &&O^{(0)} = \frac{1}{\sqrt{D}}O^{ii}, \quad  [O^{(1)}]^{ij} = \frac{1}{\sqrt{D(D-1)/2}}O^{[ij]}, \quad [O^{(2)}]^{ij} = \frac{1}{\sqrt{D(D+1)/2-1}}[O^{\{ij\}}-\frac{1}{D}O^{kk}\delta^{ij}] \,,\label{OO}
\end{eqnarray}
\end{widetext}
with $D = 3$ and
\begin{equation}
\eta^{(J)} =\frac{{\rm Sp}\, V O^{(J)} V O^{(J)} }{{\rm Sp}\,  O^{(J)}  O^{(J)}}\,.
\label{etaQ}
\end{equation}
with
\begin{equation}
V =  \left(\begin{array}{cc} 0  & \hat{k}_{+} \sigma\\
-\hat{k}_{+} \sigma& 0 \end{array} \right)
\end{equation}

In the $v$ - channel at $J=0$  the energy gap is equal to zero that leads to  the condition
\begin{equation}
 {\rm const} \, \int_{4\Delta^2}^{\Lambda^2}\sqrt{1 - \frac{4 \Delta^2}{t}} \, dt =\frac{3}{g},\label{consisthe}
\end{equation}
where $\Lambda$ is the ultraviolet cutoff.  The bosonic energy gaps $E_{u,v}^{(J)}$ are defined by the equation:
\begin{equation}
 {\rm const} \, \int_{4\Delta^2}^{\Lambda^2}\sqrt{1 - \frac{4 \Delta^2}{t}} \frac{ t  - 2 \Delta^2(1\pm  \eta^{(J)})}{t -  [E^{(J)}_{u,v}]^2}\, dt = \frac{3}{g} \label{MQhe}
\end{equation}
with the same constant as in Eq. (\ref{consisthe}). Comparing these two equations we come to

\begin{lemma}
 The energy gaps are given by
\begin{equation}
E_{u,v}^{(J)} = \sqrt{ 2 \Delta^2(1\pm  \eta^{(J)})}  \,,
\end{equation}
which  proves the Nambu sum rule for $^3$He-B:
\begin{equation}
[E_{u}^{(J)}]^2 + [E_{v}^{(J)}]^2  = 4 \Delta^2\label{NSRThe}
\end{equation}
\end{lemma}

Explicit calculation of \eqref{etaQ} gives  $\eta^{J=0} = \eta^{J=1} = 1$, and $\eta^{J=2} = \frac{1}{5}$. Thus we get immediately the result obtained in  \cite{He3B} via the direct solution of the equation ${\rm Det}\, \Bigl(g\Pi(i E) - 1\Bigr) = 0$:

\begin{enumerate}

\item{$J=0$. }

For $J=0$ there is one pair of the Nambu partners (the gapless Goldstone sound mode and the so-called pair-breaking mode  with the energy gap $E=2 \Delta$):
\begin{equation}
E^{(0)}_1 = 0,\quad E^{(0)}_2 =  2 \Delta
\end{equation}
\item{$J=1$. }

For $J=1$ there are three  pairs of Nambu partners (three gapless Goldstone modes --  spin waves and three corresponding pair-breaking modes with the energy gap $E=2 \Delta$):
\begin{equation}
E^{(1)}_1 = 0,\quad E^{(1)}_2 =  2 \Delta
\end{equation}

\item{$J=2$. }

For $J=2$ there exist five pairs -- five the so-called real squashing modes with the energy gap $E = \sqrt{2/5}\, (2\Delta)$ and correspondingly  five imaginary squashing modes with the energy gap $E=\sqrt{3/5}\, (2 \Delta)$:
  \begin{equation}
E^{(2)}_1 = \sqrt{2/5}\, (2\Delta),\quad E^{(2)}_2 =  \sqrt{3/5}\, (2\Delta) \,.
\label{He3B2}
\end{equation}
(Zeeman splitting of imaginary squashing mode in magnetic field has been observed in \cite{Lee1988},
for the latest experiments  see  \cite{Collett2012}.)

\end{enumerate}

\subsection{Nambu sum rules in $^3$He-A}


In the A - phase of $^3$He the condensate is formed in the state with $S_z=0$ and $L_z=1$ \cite{VollhardtWolfle1990}.  In the orbital sector the symmetry breaking in $^3$He-A is similar to that in the electroweak theory:
$U(1)\otimes SO_L(3)\rightarrow U_Q(1)$, where the quantum number $Q$  plays the role of the electric charge (see e.g. Ref. \cite{VolovikVachaspati1996}), while in the spin sector one has
$SO_S(3)\rightarrow SO_S(2)$.
According to \cite{BrusovPopov1980} one has
\begin{widetext}
\begin{equation}
c^{(0)}_{i \alpha}(p) = (\beta V)^{1/2} C \, \delta_{p0}(\delta_{i 1}+i\delta_{i 2}) \delta_{\alpha 3} =  (\beta V)^{1/2} C \, \delta_{p0}
\left(\begin{array}{ccc} 0 &0 & 1\\
0 & 0 & i \\
0&0&0\end{array}\right) \,,
\end{equation}
\end{widetext}
where  $C$ satisfies the gap equation
\begin{equation}
 0=\frac{1}{g}- \frac{2}{\beta V} \sum_p\frac{1-\hat{k}_3^2}{\omega^2 + v_F^2(|k|-k_F)^2+4 C^2 (1-\hat{k}_3^2)} \,.
\end{equation}

The A-phase is anisotropic. The special direction  in the orbital  space appears that is identified with the direction of the spontaneous orbital angular momentum of Cooper pairs, which is here chosen along the axis $z$.
In this phase fermions are gapless.  However, the value $\Delta(\theta) = 2 C \sqrt{1-\hat{k}_3^2} = \Delta_0 {\rm sin} \theta$ may be considered as the technical gap depending on the direction in space that enters the expressions to be considered below. (Here $\theta$ is the angle between the anisotropy axis and the direction of the momentum $k$.)

In the BCS theory of $^3$He-A, all bosonic modes  are triply degenerate. This is the consequence
of the hidden symmetry of the BCS theory applied to $^3$He-A, which in particular gives rise to 9 gapless Goldstone  modes instead of 5 modes required by symmetry breaking \cite{VolovikKhazan1982,Volovik1990}. On the language of $c_{i \alpha}$ this hidden symmetry leads to the representation of the one - loop effective action  as the sum of the three terms. Each of that terms depends on $c_{i \alpha}$ with definite value of $\alpha = 1,2,3$. The term with $c_{i 3}$ is transformed into the term with $c_{i2}$ via the substitution $c_{i 3}\rightarrow ic_{i2}$. The term with  $c_{i 2}$ is transformed into the term with $c_{i1}$ via the substitution $c_{i 2}\rightarrow c_{i1}$. Among $5$ Goldstone bosons corresponding to the
breakdown pattern $U(1)\otimes SO_L(3)\otimes SO_S(3) \rightarrow U_Q(1)\otimes SO_S(2)$ there are $u_{11} + v_{21}, u_{12}+v_{22}, u_{23} - v_{13}$ that are transformed to each other by the mentioned above transformation. Also there are the Goldstone modes $u_{33,} v_{33}$. The latter modes may be transformed by this transformation to $u_{31}, u_{32}, v_{31}, v_{32}$. Therefore, four additional gapless modes appear in weak coupling limit. Recall that in the strong coupling regime these four
modes become gapped.

The values of the energy gaps are given by the solutions of the equation ${\rm Det}\, \Bigl(g\Pi(i E) - 1\Bigr) = 0$.
Exact solutions of the given equations are presented in \cite{BrusovPopov1980}. The energy gaps are complex - valued that means that the states are not stable. (The decay into the massless fermions is possible.) However, the real parts of the energy gaps can be evaluated in the approximation, when the effective action at $k=0$ is represented as the sum of the two terms: the first term corresponds to $\omega = 0$ while the second term is proportional to $\omega^2$. Such a calculation gives the mass term for the modes of the field $c_{i\alpha}$ with the contribution due to the terms depending on higher powers of $\omega$ disregarded. This procedure gives six unpaired gapless Goldstone modes and two  pairs of modes (triply degenerated) that satisfy a version of the Nambu sum rule.
In this case the role of the square of the fermion mass is played by the angle average of the square of the anisotropic gap:
\begin{equation}
\bar{\Delta}^2 \equiv  \left<\Delta^2(\theta)\right> =\frac{2}{3}\Delta_0^2 \,.
\label{FermionMass}
\end{equation}
The Nambu pairs are the following:
\begin{enumerate}
\item{}

One  (triply degenerated) pair of bosons (the phase and amplitude collective modes in Nambu terminology) is formed by the ``electrically neutral''  ($Q=0$) massless Goldstone mode and the  ``Higgs boson''  also with $Q=0$:
\begin{equation}
E^{(Q=0)}_1 = 0,\quad E^{(Q=0)}_2 =  2 \bar{\Delta}   = \sqrt{8/3} {\Delta_0}\label{A1}
\end{equation}

\item{ }

The other  (triply degenerated) pair represents the analog of the charged Higgs bosons  in $^3$He-A with $Q=\pm 2$ (see e.g. \cite{Volovik1990}). These are the so-called clapping modes whose energies are

\begin{equation}
E^{(Q=2)}_1 = E^{(Q=-2)}_2 =  \sqrt{2} \bar{\Delta} = \sqrt{4/3} {\Delta_0} \label{Q2}
\end{equation}

\end{enumerate}

\begin{lemma} {
One can see that the spectrum of fermions and bosons in anisotropic superfluid $^3$He-A also satisfies the Nambu conjecture written in the form
\begin{equation}
E^2_1 + E^2_2 = 4 \bar{\Delta}^2
\label{ClappingModes}
\end{equation}
(for each of the two pairs listed above) with the "average fermion  gap" given by Eq. (\ref{FermionMass}).}
\end{lemma}

Alternatively these values may be obtained if in Eq. (2.16) of \cite{BrusovPopov1980} the
values of $\Delta^2(\theta)$ are substituted by their averages $\bar{\Delta}^2 \equiv  \left<\Delta^2(\theta)\right> =\frac{2}{3}\Delta_0^2$. Then, the integrals are omitted and we obtain the above listed values of the gaps.

As it was mentioned above, in the anisotropic systems in which the fermionic energy gap has zeroes, the spectrum of
massive collective modes has imaginary part due to radiation of the gapless fermions. That is why
the Nambu rule is not obeyed for the pole masses, but is obeyed for the mass parameters which are real,
since they are determined at $\omega=0$. In the systems, in which radiation is absent, such as isotropic  fully gapped superfluid $^3$He-B, the pole masses of the collective modes coincide with their mass parameters.

\subsection{Superfluid  phases in 2+1 films}

The same relations \eqref{A1} and \eqref{Q2} take place for the bosonic collective modes in the quasi two-dimensional superfluid $^3$He films. There are two possible phases in thin films,  the a-phase and the so-called planar phase (b phase in the terminology of Ref. \cite{BrusovPopov1981}). Both phases have isotropic gap $\Delta$ in the 2D case, as distinct from the 3D case where such phases are anisotropic with zeroes in the gap.

We have the effective action for the bosonic degrees of freedom Eq. (\ref{hydrodynamic}), Eq. (\ref{matrix}) with the $2\times 3$ matrix $c_{i \alpha}$. The following two forms of these matrices correspond to the  a - and b - phases \cite{BrusovPopov1981}:
 \begin{eqnarray}
c^{(0)}_{i \alpha}(p)& =& (\beta V)^{1/2} C \, \delta_{p0}
\left(\begin{array}{ccc} 1 &0 & 0\\
i & 0 & 0 \end{array}\right)\, (a-phase) \nonumber\\
c^{(0)}_{i \alpha}(p) &=& (\beta V)^{1/2} C \, \delta_{p0}
\left(\begin{array}{ccc} 1 &0 & 0\\
0 & 1 & 0 \end{array}\right)\, (b-phase) \,.
\end{eqnarray}

Let us consider the second possibility (the planar b - phase).  We have the symmetry breaking pattern
$SO(2) \otimes SO(3) \otimes U(1)\rightarrow SO(2)$. Correspondingly, there are four gapless Goldstone modes. Among them there are $u_{13}$ and $u_{23}$ modes. Modes $v_{13}$ and $v_{23}$ are their partners with the energy gaps $2 \Delta$. The analysis is similar to that of the s - wave superconductor.

As for the modes $u_{ij}, v_{ij}$ with $i,j = 1,2$, the analysis is similar to that of the
$^3$He-B phase. The spectral densities $\rho_{u,v}$ differ from those of Eq. (\ref{RHO}) by the kinematic factor $\sqrt{1/t}$ instead of $\sqrt{1-4 \Delta^2/t}$.  Next, we substitute $D=2$ to Eq. (\ref{OO}), and get
\begin{equation}
E_{u,v}^{(J)} = \sqrt{ 2 \Delta^2(1\pm  \eta^{(J)})}  \,, J = 0,1,2
\end{equation}
Direct calculation of \eqref{etaQ} gives  $\eta^{J=0} = - \eta^{J=1} = 1$, and $\eta^{J=2} = 0$. (In this case $J$ is not the total momentum of the Cooper pair. )

\begin{lemma}
{\it The resulting spectrum in b - phase is}
\begin{eqnarray}
E_{u,v}^{(0)} &=&2 \Delta, 0; \, E_{u,v}^{(1)} = 0, 2 \Delta; \, {\rm and}\nonumber\\~~~~ E_{1,u,v}^{(2)}&=& \sqrt{2}\Delta; \,E_{2,u,v}^{(2)}= \sqrt{2}\Delta\, .
\label{2Db}
\end{eqnarray}
\end{lemma}

In the $a$ phase the symmetry breaking is
$SO(2) \otimes SO(3) \otimes U(1) \rightarrow U(1)_Q\otimes SO(2)$ with three Goldstone modes.
Acting as above, for the b - phase (in this case the $u$ and $v$ modes are mixed unlike the b - phase) or, applying the results of Ref. \cite{BrusovPopov1981}, one obtains

\begin{lemma}
{ These modes of the a - phase form two pairs of Nambu partners (triply degenerated), with $Q=0$ and $|Q|=2$:}
\begin{eqnarray}
E_1^{(Q=0)} &=&0~~,~~ E_2^{(Q=0)} = 2 \Delta~~~~{\rm and}\nonumber\\~~~~ E^{(Q=+2)}&= &\sqrt{2}\Delta~~,~ E^{(Q=-2)}= \sqrt{2}\Delta  \,.
\label{2D}
\end{eqnarray}
\end{lemma}

Note that since masses of $Q=+2$ and $Q=-2$ modes are equal,
the Nambu sum rule necessarily leads
to the definite value of the masses of the ``charged'' Higgs bosons.

It is worth mentioning that, in principle,  the derivation of the energy gaps for the $a$ phase with minor modifications may be applied also for the evaluation of the real parts of the energy gaps of the $3D$  A - phase. In such calculations dealing with the equations that are the analogues of Eq. \eqref{etaQ} we need to substitute the angle averaged fermionic gap  \eqref{FermionMass}.

Because of the common symmetry breaking scheme in the electroweak theory and in $^3$He-A we consider the listed above energy gaps as an indication of the existence of the Higgs boson with mass
\begin{equation}
M_H=\sqrt{2} M_T  \,.
\label{2D}
\end{equation}
This mass is about 245 GeV, which is roughly twice the mass of the lowest energy Higgs boson.

\section{Nambu sum rules in the relativistic models of top quark condensation}


\subsection{Effective NJL model for the dynamical electroweak symmetry breaking}

\label{NJL1loop}

In this section we consider the extended NJL model of top - quark condensation. This model was suggested in \cite{Miransky} by Miransky and coauthors and generalizes the more simple models (see, for example, \cite{topcolor1,2HMC,2HMMiransky}).  It  includes all 6 quarks.  At the present moment we do not wish to define the realistic theory aimed to explain DEWSB and the formation of fermion masses. Our objective is to demonstrate how the Nambu sum rule (probably, in the modified form) may appear in the relativistic models of general kind.

The most general form of the four - fermion action (for the model with 6 quarks) has the form
\begin{widetext}
\begin{equation}
S = \int d^4x \Bigl(\bar{\chi}[ i \nabla \gamma ]\chi +  g (\bar{\chi}_{\alpha A,L} \chi^{\beta B}_R)(\bar{\chi}_{\bar{{\beta}} \bar{B}, R} {\chi}^{\bar{\alpha} \bar{A}}_{L}) [YY^+]_{\bar{\alpha}\bar{A} \beta B}^{\alpha A \bar{\beta} \bar{B}}+g (\bar{\chi}_{\alpha A,L} \chi^{\beta B}_R)(\bar{\chi}_{\bar{{\beta}} \bar{B}, L} {\chi}^{\bar{\alpha} \bar{A}}_{R}) [WW^+]_{\bar{\alpha}\bar{A} \beta B}^{\alpha A \bar{\beta} \bar{B}}\Bigr) \label{Stopcolor_}
\end{equation}
\end{widetext}
Here $\chi_{\alpha A}^T = (u,d); (c,s); (t,b)$ is the set of the doublets with the generation index $\alpha$. Tensors $W,Y$ contain coupling constants.
We consider the particular case of this model, when $W=0$, while tensor $Y$ is factorized:
\begin{equation}
Y_{\bar{\alpha}\bar{A} \beta B}^{\alpha A \bar{\beta} \bar{B}}=L_{\bar{\alpha}}^{\alpha} R_{\beta}^{\bar{\beta}} I_B^{\bar{B}}\delta_{\bar{A}}^A,
\end{equation}
where $L,R,I$ are Hermitian. Here it is taken into account that the electroweak symmetry has to be preserved.
The given four - fermion action approximates the microscopic theory. We suppose that this unknown microscopic theory has the approximate $U(2\times 3)_L\otimes U(2\times3)_R$ symmetry that is broken softly down to $U(2)_L\otimes U(2)_L\otimes U(2)_L\otimes U(1)_R \otimes ... \otimes U(1)_R$. In the zeroth order approximation all eigenvalues of matrices $L,R,I$ are equal to each other. In the next approximation this symmetry is violated softly, and the eigenvalues of $L,R,I$ receive small corrections.

\begin{remark}
The field theory with action Eq. (\ref{Stopcolor_}) is not renormalizable. The ultraviolet divergences become stronger and stronger when the number of loops is increased. Therefore, Eq. (\ref{Stopcolor_}) describes the phenomenological low energy theory. It has sense only when a finite ultraviolet cutoff $\Lambda$ is specified. The predictions of this model become independent of the regularization scheme only for the characteristic energies $\cal E$ of the processes much smaller than $\Lambda$. The physical quantities may, in principle, be evaluated using the $1/N_C$ perturbation theory. The leading terms in the expansion in the powers of $1/N_C$ correspond to the mean field approximation and are limited to the one - loop diagrams. Most of the practical calculations in the NJL - like models are performed in this approximation.  (For the review of the calculations in NJL -  approximation applied to the models of top quark condensation see, for example, \cite{cveticreview} and references therein.) The next to leading (NTL) order approximation corresponds to the number of fermion loops larger than one, or, equivalently, to the appearance of meson loops \cite{cvetic}. It has been shown that the NTL contributions to various dimensional quantities are small compared to the leading order $1/N_C$ results only for $M_t \sim \Lambda$. For $5 M_t \leq  \Lambda$ due to the NTL contribution the Higgs condensate vanishes \cite{cvetic}. Rough consideration of the NTL (and higher) contributions to the scalar meson (Higgs boson) masses gives the values of the order of $\Lambda$ unless these mesons are protected from being massive by symmetry. (For example, Goldstone theorem protects the Goldstone bosons from masses if the chiral symmetry is broken spontaneously.)
\end{remark}

The model considered in this paper corresponds to the condition $\Lambda \gg M_T$. Correspondingly, the one - loop results in the complete field model with  action  Eq. (\ref{Stopcolor_}) are not valid neither for the fermion masses nor for the masses of the bosonic excitations if one considers the model nonperturbatively or sums higher loop contributions. The one - loop prediction of the appearance of the dynamical chiral symmetry breaking may be incorrect as well. However, {\it we suggest to consider action Eq. (\ref{Stopcolor_}) as the action of the effective theory, in which only the leading $1/N_C$ (i.e. one - loop) contribution is taken into account while the higher loop corrections are to be disregarded}. Strictly speaking, this means that the quantum field theory considered here is not the one with the action of Eq. (\ref{Stopcolor_}). Namely, the complete action of this theory is to contain additional terms that cancel the quadratically divergent contributions to the fermion and meson masses. For example, the dominant contributions to the meson masses of the diagrams with $K$ four - fermion vertexes  are $\delta M_H^2 \sim g^K \Lambda^{2K-2}$. Assuming that the one - loop gap equation works we get $g \sim \frac{1}{\Lambda^2}$. As a result the higher loop contributions to the Higgs boson masses are $\delta M_H^2 \sim \Lambda^{2}$ like in the Weinberg - Salam model, where the loop corrections give quadratically divergent contributions to the Higgs boson masses. This results in the so - called hierarchy problem. One can see, therefore, that the hierarchy problem of the Standard model is reflected by the effective theory with the action of Eq. (\ref{Stopcolor_}).
In the same way the linear divergent contributions to the fermion masses appear due to the higher loops.
The mentioned above additional terms to be added to this action are to cancel these divergent contributions to meson and fermion masses. If so, the higher loop contributions both to the fermion and boson masses are suppressed by the powers of $\frac{\cal E}{\Lambda}$, where $\cal E$ is the characteristic energy of the considered processes.

At a first glance it is difficult to imagine the reasonable mechanism for the appearance of such terms.  However, there exists the theory, where in the similar situation such terms do exist. Namely, in quantum hydrodynamics \cite{quantum_hydro} there formally exist the divergent contributions to various quantities  (say, to vacuum energy) due to the quantized sound waves. The quantum hydrodynamics is to be considered as a theory with finite cutoff $\Lambda$. The loop divergences in the vacuum energy are to be subtracted just like we do for the case of the NJL model of this section. In hydrodynamics the explanation of such a subtraction is that the microscopic theory to which the hydrodynamics is an approximation works both at the energies smaller and larger than $\Lambda$, and this microscopic theory contains the contributions from the energies larger than $\Lambda$. These contributions exactly cancel the divergences appeared in the low energy effective theory. This exact cancellation occurs due to the thermodynamical stability of vacuum. In \cite{hydro_gravity} it was suggested that a similar pattern may provide the mechanism for the cancellation of the divergent contributions to vacuum energy in quantum gravity and divergent contributions to the Higgs boson mass in the Standard Model. Namely, the contributions of the trans - $\Lambda$ degrees of freedom to the given quantities exactly cancel the divergent contributions of the effective low energy theories  (correspondingly, of gravity and of the Weinberg - Salam model). {\it We suppose, that in our case of the NJL model the contributions of the trans - $\Lambda$ degrees of freedom cancel the dominant divergences in the bosonic and fermionic masses leaving us with the one - loop approximation as an effective tool for the evaluation of physical quantities.}

\subsection{One - loop effective action for the bosonic modes}

Via the suitable redefinition of the fermions we make matrices $L,R,I$ diagonal.  We denote
\begin{widetext}
\begin{eqnarray}
&& L = {\rm diag}(L_{ud},L_{cs},L_{tb}); \quad R = {\rm diag}(R_{ud},R_{cs},R_{tb}); \quad I = {\rm diag}(I_{up},I_{down});\nonumber\\
&& y_u  = L_{ud} R_{ud} I_{up}, \quad y_d = L_{ud} R_{ud} I_{down}, \quad y_c  = L_{cs} R_{cs} I_{up},\nonumber\\&& y_s = L_{cs} R_{cs} I_{down}, \quad y_t = L_{tb} R_{tb} I_{up}, \quad y_b = L_{tb} R_{tb} I_{down} \nonumber\\ &&
y_{ud}  = L_{ud} R_{ud} I_{down}, \quad y_{du} = L_{ud} R_{ud} I_{up},\quad   y_{uc}  = L_{ud} R_{cs} I_{up},\nonumber\\&& y_{cu} = L_{cs} R_{ud} I_{up},\quad y_{us} = L_{ud} R_{cs} I_{down}, \quad y_{su} = L_{cs} R_{ud} I_{up}, ...\nonumber\\
&& ...
\end{eqnarray}
\end{widetext}

We can rescale the coupling constants in such a way that
\begin{equation}
y_q = 1 + \delta y_q, \quad y_{q_1q_2}=1+\delta y_{q_1q_2}
\end{equation}
where $|\delta y_q|, |\delta y_{q_1q_2}| << 1$. The values of $\delta y_q, \delta y_{q_1 q_2}$ satisfy
\begin{equation}
\delta y_{q_1q_2}+\delta y_{q_1q_2} =
\delta y_{q_1}+\delta y_{q_2}
\end{equation}

The whole symmetry of Eq. (\ref{Stopcolor_}) is $U_{L,1}(2)\otimes U(2)_{L,2} \otimes U(2)_{L, 3} \otimes U(1)_u \otimes ... \otimes U(1)_b$.
As in the previous sections we introduce the bosonic variable $c^{\beta B}_{\alpha A}$  and insert into the functional integral the expression $1 \sim \int D \bar{c} D c\,  {\rm exp}(-\frac{i}{g}\sum_{p} \bar{c}^{\beta B}_{\alpha A}(p) c^{\alpha A}_{\beta B}(p))$. We arrive at the action

\begin{equation}
S_{eff} = -\frac{1}{g} \sum_{p} \bar{c}^{\beta B}_{\alpha A}(p) c^{\alpha A}_{\beta B}(p) + {\rm log} \, {\rm Det} M(\bar{c},c), \label{seffM_}
\end{equation}
where
\begin{eqnarray}
&&\bar{\chi}_{p_1} M(\bar{c},c) \chi_{p_2} = \bar{\chi}_{p_1}\hat{p} \gamma {\chi}_{p_2} \delta_{p_1 p_2} -
 \\&&\frac{1}{(\beta V)^{1/2}} ( Y_{\bar{\alpha} \bar{A} \beta B}^{\alpha A \bar{\beta} \bar{B}} c_{\bar{\beta} \bar{B}}^{\bar{\alpha} \bar{A}}(p_1-p_2)\bar{\chi}_{p_1,\alpha, A, L}\chi^{\beta B}_{p_2,R}  + h.c.)\nonumber
\end{eqnarray}

The equation that defines the vacua of the model is
\begin{equation}
\frac{\delta}{\delta c^{\alpha A}_{\beta B}} S_{eff} = 0\label{GAP}
\end{equation}

The solution of this equation corresponds to the stable vacuum if at the vacuum value of $c$ we have ${\rm Det} \frac{\delta^2}{\delta c^{\alpha A}_{\beta}\delta c^{\alpha^{\prime} A^{\prime} }_{\beta^{\prime}}} S_{eff} \ge 0$. This occurs if the masses of all Higgs bosons are real.
Suppose that the vacuum is CP invariant
and the vacuum  value of $c$ is equal to
 \begin{equation}
 c^{(0) \alpha A}_{\beta B}(p) = (\beta V)^{1/2} C^{ \alpha A}_{\beta B}\delta_{p0} \in R,\label{CCPI_}
\end{equation}
We also require that the mass matrix for the fermions ${\bf M}_{\beta B}^{\alpha A }  = Y_{\bar{\alpha} \bar{A} \beta B}^{\alpha A \bar{\beta} \bar{B}} C_{\bar{\beta} \bar{B}}^{\bar{\alpha} \bar{A}}$ is Hermitian, then Eq. (\ref{GAP}) in one loop approximation has the form:
\begin{widetext}
\begin{equation}
C^{ \alpha A}_{\beta B} = \frac{i g}{2 (2\pi)^4} Y_{\bar{\alpha} \bar{A} \beta B}^{\alpha A \bar{\beta} \bar{B}}  \int {\rm Tr}\,\Bigl( \frac{1}{{l} \gamma  -
  {\bf M} }\Bigr)^{\bar{\alpha}\bar{A}}_{\bar{\beta}\bar{B}}\, d^4 l  = \frac{2ig N_C}{(2\pi)^4} Y_{\bar{\alpha} \bar{A} \beta B}^{\alpha A \bar{\beta} \bar{B}} \int \,\Bigl( \frac{{\bf M}}{l^2  -
  {\bf M}^2  }\Bigr)^{\bar{\alpha}\bar{A}}_{\bar{\beta}\bar{B}}\, d^4 l\,
\end{equation}
\end{widetext}
Here $N_C=3$ is the number of colors.

This equation has many different solutions that correspond to different vacua. We consider here only the case, when
the matrices $C$ and $\bf M$ are diagonal, so that there exist the condensates $\langle q\bar{q} \rangle$ and nonzero masses for all quarks. We also imply that the t - quark mass and the t - quark condensate dominate. The quark masses  $M_q = y_q C_q$ satisfy the equations
\begin{eqnarray}
&&0= \frac{1}{g N_C}  -  \frac{2i}{(2\pi)^4} y_q^2  \int \,\Bigl( \frac{1}{l^2  -
 M_q^2  }\Bigr)\, d^4 l \nonumber\\ &&= \frac{1}{g N_C} - \frac{y_q^2}{8\pi^2} (\Lambda^2 - M_q^2 \, {\rm log}\,\frac{\Lambda^2}{M_q^2}), \label{gap__}
\end{eqnarray}
where $\Lambda$ is the ultraviolet cutoff. If we set
\begin{equation}
g = \frac{8\pi^2}{ \Lambda^2 N_C}
\end{equation}
then at $\Lambda >> M_q$ from the gap equations it follows that
\begin{equation}
\frac{M_q^2}{\Lambda^2}{\rm log} \frac{\Lambda^2}{M_q^2} = 2 \delta y_q
\end{equation}

The given vacuum  does not exhaust all possible vacua of the model.
Further we shall imply that the external conditions (and the values of couplings) are such that the given vacuum wins the competition between all possible vacua.

The symmetry breaking pattern is $U_{L,1}(2)\otimes U(2)_{L,2} \otimes U(2)_{L, 3} \otimes U(1)_u \otimes ... \otimes U(1)_b \rightarrow U(1)_u\otimes ...\otimes U(1)_t\otimes U(1)_b$. Therefore, we expect the appearance of $12$ Goldstone bosons. Only three of them are to be eaten by Z and W bosons. In order to make the other modes massive the gauge fields may be added that become massive due to the symmetry breaking and absorb the mentioned extra Goldstone bosons.

\subsection{Higgs bosons masses}

The bosonic masses can be calculated as follows.
As in Section 2 at $k = 0$ the polarization operator can be represented as
\begin{equation}
\Pi(\omega) = \int_{0}^{\infty} d z \frac{\rho(z)}{z+\omega^2}, \label{disp}
\end{equation}
with the spectral function $\rho$.

In the scalar/pseudoscalar  $q\bar{q}$ channel we have ($(\sqrt{t},0) = p_+ + p_- ; \, p_{\pm}^2=M_q^2$):
\begin{widetext}
\begin{eqnarray}
\rho^{S}_{q\bar{q}}(t) & = & \frac{1}{32\pi^2}\theta(t-4 M_q^2) \, \sqrt{1 - \frac{4 M_q^2}{t}} \, {\rm Sp}\, (\gamma p_- +M_q) (\gamma p_+ -M_q) =   \frac{1}{16\pi^2}\sqrt{1 - \frac{4 M_q^2}{t}} \, (t  - 4 M_q^2) \theta(t-4 M_q^2) \,\nonumber\\
\rho^{P}_{q\bar{q}}(t) & = &   \frac{1}{32\pi^2}\theta(t-4 M_q^2) \,\sqrt{1 - \frac{16 M_q^2}{t}} \, {\rm Sp}\,i \gamma^5 (\gamma p_- +M_q)i \gamma^5 (\gamma p_+ -M_q) =   \frac{1}{16\pi^2} \sqrt{1 - \frac{4 M_q^2}{t}} \, t \theta(t-4 M_q^2)\label{DISP}
\end{eqnarray}
\end{widetext}

Integrals in Eq. (\ref{disp}) are ultraviolet divergent. The regularization may be introduced in such a way that the upper limit in each integral is substituted by the finite cutoff (that may depend on the channel). Next, the $(q\bar{q})$ condensate provides the symmetry breaking. There should be  Goldstone bosons corresponding to the broken symmetry. This  provides P excitation in $q\bar{q}$ channel is massless (the corresponding bilinear appears via the application of the generator of the broken symmetry to $(q\bar{q})$). Then, we have $\Pi_{q\bar{q}}^{P}(0) = \Pi_{q\bar{q}}^{S}(2iM_q)$ that means that the massive scalar excitation appears with mass $2 M_T$. The same result can be obtained in the
neutral channels $q \bar{q}$ via the direct calculation of the polarization operator:
\begin{widetext}
\begin{eqnarray}
\frac{1}{g N_C}-\Pi^S_{q\bar{q}}(i E) & = &\frac{1}{g N_C} + \frac{i  y^2_q}{2(2\pi)^4} \, \int d^4 l \, {\rm Sp}\,\frac{1}{l \gamma -M_q }\, \frac{1}{(p-l)\gamma-M_q}\nonumber\\
&=& {(p^2-4M_q^2)  y^2_q } \, I(M_q,M_q,p) \nonumber\\
\frac{1}{g N_C}-\Pi^P_{q\bar{q}}(i E) & = &\frac{1}{g N_C} + \frac{i  y^2_q}{2(2\pi)^4} \, \int d^4 l \, {\rm Sp}\,i \gamma^5\frac{1}{l \gamma -M_q }\,i \gamma^5 \frac{1}{(p-l)\gamma-M_q}\nonumber\\
&=& {(p^2)  y^2_q } \, I(M_q,M_q,p),\nonumber\\&&
I(m_1,m_2,p) = \frac{i}{(2\pi)^4} \, \int d^4 l \, \frac{1}{(l^2- m_1^2)[(p-l)^2-m_2^2]}\label{MQQ}
\end{eqnarray}
\end{widetext}
Here the gap equation is used.
We get
\begin{equation}
M^P_{q\bar{q}} = 0; \quad M^S_{q\bar{q}} = 2 M_q\label{NSRS}
\end{equation}
for  $q = u,d,c,s,t,b$.

\begin{remark} {The calculations of the bosonic spectrum in NJL models suffer from the ambiguity appeared when the shift of the variable is performed in the integral $\int \frac{d^4l}{(l-p)^2-m^2} \rightarrow \int \frac{d^4l}{l^2-m^2}$. In fact this change of variables is not rigorous and results in the appearance of the new surface terms. This is a very - well known problem of the NJL models. (See, for example, \cite{AMBIG}).) The resulting contributions  due to the surface terms were evaluated in Eq. (37) of \cite{AMBIG}. From \cite{AMBIG} it follows that in the limit of large cutoff $\Lambda$  the extra contributions to   Eq. (\ref{NSRS}) vanish. It should be stressed that this problem does not appear in the dimensional regularization. In the lattice regularization momentum space is the $4D$ torus, and the shift of the integration variable in the integrals like $\int \frac{d^4l}{(l-p)^2-m^2}$ is performed easily. Instead, in both cases their own problems appear like the fermion doubling problem. However, in the approach through the dispersion relation Eq. (\ref{DISP}) this problem does not appear while the final result is again $2M_q$. This justifies indirectly the shift of the variable in Eq.(\ref{MQQ}) $l-p\rightarrow l$.  Anyway, the phenomenologically justified  value of the NJL calculations in QCD and in the other models allows us to disregard the mentioned problem and to perform the shift of the integration variables in the calculation of the bosonic spectrum.}
\end{remark}

In each channel that includes two different quarks $q_1, {q_2}$ the polarization operator is a $2\times 2$ complex matrix $\cal P$.  For its components we have
\begin{widetext}
\begin{eqnarray}
\frac{1}{g N_C}-\Pi_{q_{1,L}\bar{q}_{2,R}}^{q_{1,L}\bar{q}_{2,R}}(i E) & = & \frac{1}{g N_C} + \frac{i  y_{q_1q_2}^2}{4(2\pi)^4} \, \int d^4 l \, {\rm Sp}\,\frac{1}{l \gamma -M_{q_1} }\, (1\pm \gamma^5)\frac{1}{(p-l)\gamma-M_{q_2}}(1\mp \gamma^5)\nonumber\\
&=& {(p^2-M_{q_1}^2-M_{q_2}^2) y_{q_1q_2}^2 }\, I(M_{q_1},M_{q_2},p) - \frac{i  y_{q_1q_2}^2}{(2\pi)^4} \, \int d^4 l \, \frac{1}{l^2 -M_{q_1}^2 } - \frac{i  y_{q_1q_2}^2}{(2\pi)^4} \, \int d^4 l \, \frac{1}{l^2 -M_{q_2}^2 } \nonumber\\ &&+  (1+\zeta_{q_1 q_2})\frac{ i  y_{q_1q_2}^2}{(2\pi)^4} \, \int d^4 l \, \frac{1}{l^2 -M_{q_1}^2 } +  (1-\zeta_{q_1 q_2}) \frac{ i  y_{q_1q_2}^2}{(2\pi)^4} \, \int d^4 l \, \frac{1}{l^2 -M_{q_2}^2 } \nonumber\\
&=& {(p^2-M_{q_1}^2-M_{q_2}^2) y_{q_1q_2}^2 }\, I(M_{q_1},M_{q_2},p)     + \zeta_{q_1 q_2} y_{q_1q_2}^2   J(M_{q_1},M_{q_2}), \, {\rm where}\nonumber\\
&&\frac{1}{y^2_{q_1q_2}} = \frac{1+\zeta_{q_1 q_2}}{2 y^2_{q_1}}+\frac{1-\zeta_{q_1 q_2}}{2 y^2_{q_2}}\label{Ut_}
\end{eqnarray}
\end{widetext}
For the cross - terms:
\begin{widetext}
\begin{eqnarray}
-\Pi_{q_{1,L}\bar{q}_{2,R}}^{q_{2,L}\bar{q}_{1,R}}(i E) & = &  \frac{i  y_{q_1q_2} y_{q_2q_1}}{4(2\pi)^4} \, \int d^4 l \, {\rm Sp}\,\frac{1}{l \gamma -M_{q_1} }\, (1\pm \gamma^5)\frac{1}{(p-l)\gamma-M_{q_2}}(1\pm \gamma^5)\nonumber\\
&=& {2 M_{q_1} M_{q_2} y_{q_1q_2}y_{q_2q_1}}\, I(M_{q_1},M_{q_2},p) \label{Utc_}
\end{eqnarray}
\end{widetext}

At $\Lambda >> M_{q_1} > M_{q_2}$ we get $J(M_{q_1},M_{q_2})\approx [M_{q_1}^2 - M_{q_2}^2] I(M_{q_1},M_{q_2},p) \approx \frac{M_{q_1}^2 - M_{q_2}^2}{16 \pi^2} \,{\rm log}\Lambda^2/M_{q_1}^2$ (here $\Lambda$ is the ultraviolet cutoff of the NJL model).
Therefore
\begin{widetext}
\begin{eqnarray}
\frac{1}{g N_C}-{\cal P}_{q_1q_2}(i E) & = &  \left(\begin{array}{cc}[E^2-M_{q_1}^2(1-\zeta_{q_1q_2}) - M_{q_2}^2(1+\zeta_{q_1q_2})]y_{q_1q_2}^2 & 2 M_{q_1} M_{q_2} y_{q_1q_2}y_{q_2q_1}\\2M_{q_1}M_{q_2}y_{q_1q_2}y_{q_2q_1}& [E^2-M_{q_1}^2 (1+\zeta_{q_2q_1})- M_{q_2}^2(1-\zeta_{q_2q_1})]y_{q_2q_1}^2
 \end{array}\right) \nonumber\\&&\times \frac{1}{16 \pi^2} \,{\rm log}\Lambda^2/M_{q_1}^2
\end{eqnarray}
\end{widetext}

Each state is doubly degenerate (we mark the corresponding states by $+$ or $-$).
We come to

\begin{lemma}
{ In the considered toy model we have two excitations in each $q\bar{q}$ channel for $q = u,d,c,s,t,b$ :
\begin{equation}
M^P_{q\bar{q}} = 0; \quad M^S_{q\bar{q}} = 2 M_q
\end{equation}
 and four excitations (i.e. two doubly degenerated excitations) in each $q_1\bar{q}_2$ channel (for  $ q_1\ne q_2 = u,d,c,s,t,b$) with masses:
\begin{eqnarray}
M_{q_1 {q}_2}^2 &= &M_{q_1}^2 + M_{q_2}^2  \\ &&\pm \sqrt{(M_{q_2}^2 - M_{q_1}^2)^2 \zeta_{q_1q_2}^2+ 4 M_{q_1}^2M_{q_2}^2},\nonumber
\end{eqnarray}
where
 $\zeta_{q_1 q_2}$ are given by \begin{equation}
 \zeta_{q_1 q_2} = \frac{2 \delta y_{q_1q_2}- \delta y_{q_2} - \delta y_{q_1}}{\delta y_{q_2}-\delta y_{q_1}} =
\zeta_{q_2 q_1}
 \end{equation}

The Nambu sum rule has the form}
\begin{widetext}
\begin{eqnarray}
&& [M^{+}_{q_1\bar{q}_2}]^2  + [M^{-}_{q_1\bar{q}_2}]^2 + [M^{+}_{q_2\bar{q}_1}]^2  + [M^{-}_{q_2\bar{q}_1}]^2\approx 4 [M_{q_1}^2 + M_{q_2}^2], \quad q_1\ne q_2 \nonumber\\
&& [M^{P}_{q\bar{q}}]^2  + [M^{S}_{q\bar{q}}]^2 \approx 4 M_{q}^2
\end{eqnarray}
\end{widetext}
\end{lemma}

\begin{remark} { At $|\zeta_{q_1q_2}|<1 $ all considered bosonic masses are real, there are no tachions, which means that the vacuum is stable.}
\end{remark}

Since the top quark mass is much larger than the other fermion masses,
the Nambu sum rule in the form of Eq. (\ref{NSR}) with the top quark mass in the r.h.s. holds in all channels that include the top quark. The other boson masses are much smaller.

\vspace{3mm}
\begin{remark}
{ Among the mentioned Higgs bosons there are 12 Goldstone bosons that are exactly massless (in the channels $t(1\pm \gamma^5)\bar b, t \gamma^5\bar{t},  c(1\pm\gamma^5)\bar{s}, c\gamma^5\bar{c}, u(1\pm \gamma^5)\bar{d}, u\gamma^5\bar{u},b\gamma^5\bar{b}, s\gamma^5\bar{s}, d\gamma^5\bar{d}$). There are Higgs bosons with the masses of the order of the t-quark mass ($ t(1\pm \gamma^5)\bar b, t \bar{t},  t(1\pm\gamma^5)\bar{s}, t\gamma^5\bar{c}, t(1\pm \gamma^5)\bar{d}, t\gamma^5\bar{u}$). The other Higgs bosons have masses much smaller than the t - quark mass.}
\end{remark}

As it was mentioned above, the simplest relativistic models of the kind discussed in this section were considered in \cite{NJL,topcolor1,2HMC}.  In these models the neutral Higgs bosons have masses $0$ or $2M_T$.  However, the model considered in   \cite{2HMC} has the charged Higgs bosons with masses $\approx \sqrt{2}M_T$. Actually, our derivation of the masses in $t\bar{q}$ channels is similar to that of \cite{2HMC} for the charged Higgs bosons.

A further generalization of the model of \cite{2HMC} was considered in  \cite{2HMMiransky}, where three scalar Higgs doublets are to be introduced, the fourth generation of quarks with large masses is involved. In this model there are two charged scalar Higgs modes with masses $M^{H^{\pm}}_1, M^{H^{\pm}}_2$,  and two pseudoscalar modes with masses $M^A_1, M^A_2$ that satisfy the following relation $2 \sum_i \Bigl([M^{H^{\pm}}_i]^2  - [M^{A}_i]^2 \Bigr)\approx 4 M_T^2$.

\subsection{Nambu sum rules in dense QCD}

Among the other relativistic systems, where the analogues of the Nambu sum rules were observed, we would like to mention QCD in the presence of finite chemical potential. First, let us notice the normal phase with the broken chiral symmetry (both T and $\mu$ are small compared to the QCD scale $\Lambda_{QCD}$). We already mentioned in the introduction, that in this phase the NJL approximation leads to the Nambu sum rule in the trivial form $M_{\sigma} = 2 M_{quark}$.

In dense QCD with $\mu > \Lambda_{QCD}$ there may appear several phases with different diquark condensates. Among them there is, for example, the color - flavor locking phase (CFL). In the phenomenological models of this phase the three quarks $u,d,s$ are supposed to be massless. The condensate is formed  \cite{CFL0, CFL}
\begin{equation}
\langle [\psi^{i}_{\alpha}]^t i \gamma^2\gamma^0 \gamma^5 \psi^j_{\beta} \rangle \sim \Phi^I_J \, \epsilon_{\alpha \beta J} \epsilon^{ijI} \sim  (\beta V)^{1/2} C \,  \, \epsilon_{\alpha \beta I} \epsilon^{ijI}
\end{equation}

There are $18$ scalar fluctuations of $\Phi$ around this condensate (there are also $18$ pseudo - scalar fluctuations with the same masses \cite{diquarks}). The symmetry breaking pattern is $SU(3)_L \otimes SU(3)_R \otimes SU(3)_F\otimes U(1)_A\otimes U(1)_B \rightarrow SU(3)_{CF}$. That's why there are $9+9$ massless Goldstone modes.  Among the remaining $9+9$ Higgs modes there are two octets of the traceless modes and two singlet  trace modes.
Correspondingly, the quark excitations also form singlets and octets. The singlet fermionic gap $\Delta_{1}$ is twice larger than the octet fermionic gap $\Delta_{8}$ (see Sect. 5.1.2. of \cite{CFL}).
Applying the technique similar to that of we developed for the consideration of $^3$He-B we get the scalar singlet and octet masses $M_1 = 2 \Delta_1, M_8 = 2 \Delta_8$.
This may also be derived from the results presented in \cite{m1,m1m8}.
Thus, for the CFL phase of the color superconductor we have the Nambu sum rules in the trivial form.

\subsection{Nambu sum rule and Veltman identity}

The condition for the cancellation of the quadratic divergences in the mass of the Higgs boson was discussed in a number of papers (see, for example, \cite{Alberghi2008,FrolovFursaev1998,FFZ2003,Zeldovich1968,veltman}
and references therein).  In the case of the single Higgs boson and in the absence of the gauge fields this condition reads:
$3  M_{H}^2  = 4 \sum_f M^2_f$.
Here in the l.h.s.  the scalar boson mass stands while in the r.h.s. the sum is over the fermions. If the model contains only triply degenerated quarks, this relation is reduced to
\begin{equation}
M_{H}^2  = 4 \sum_f M^2_f \label{VELTMAN}
\end{equation}
There is an obvious analogy between this identity and the Nambu sum rule Eq. (\ref{NSR}). Let us consider this analogy in more details.

In this particular case the bare action for the model that involves the scalar doublet $c$ and the fermion fields $\chi^{q}$ has the form:
\begin{widetext}
\begin{eqnarray}
S_{eff} & = &  \frac{1}{2}\sum_{p} \bar{c}(p) [p^2 + \frac{{M_H}^2}{2} ]c(p) -
\frac{\lambda^2}{8} \sum_{p_1-p_2=p_3-p_4} {\bar{c}}(p_1) c(p_2){\bar{c}}(p_4) c(p_3) \nonumber\\&&+
 \sum_{p} \bar{\chi}_{p}{p} \gamma {\chi}_{p}  -  \sum_{q,p=p_1-p_2}
  ( y_{q   } c(p)\bar{\chi}_{p_1,q, L}\chi^{q }_{p_2,R}  + h.c.)
\end{eqnarray}
\end{widetext}

  ${M}_H$ is equal to the bare mass of the scalar.  Masses of the fermions are related to this value as $M_{q} = \frac{y_{q} M_H}{\lambda} $ The origin of Eq. (\ref{VELTMAN}) is in the expression for the one - loop correction to the Higgs mass
\begin{widetext}
\begin{eqnarray}
\Pi^{(2)} & \sim & \frac{ i 3 \lambda^2}{2(2\pi)^4} \, \int d^4 l \, \frac{1}{l^2 -M_H^2 } + \frac{ i 3 \lambda^2}{2(2\pi)^4} \, \int d^4 l \, \frac{1}{l^2} + \sum_q\frac{i N_C y_{q}^2}{(2\pi)^4} \, \int d^4 l \, {\rm Sp}\,\frac{1}{l \gamma -M_{q} }\, \frac{1}{(p-l)\gamma-M_{q}}|_{p=0}\nonumber\\
&=& 2 N_C\sum_q{(-4M_{q}^2) y_{q}^2 }\, I(M_{q},M_{q},0) - \lambda^2 \sum_q\frac{4 N_C i  M_{q}^2}{M_H^2(2\pi)^4} \, \int d^4 l \, \frac{1}{l^2 -M_{q}^2 } +  \frac{3 i  \lambda^2}{2(2\pi)^4} \, \int d^4 l \, \frac{1}{l^2 -M_{H}^2 } \nonumber\\
&& +  \frac{3 i  \lambda^2}{2(2\pi)^4} \, \int d^4 l \, \frac{1}{l^2  } \approx  \frac{\Lambda^2}{16\pi^2}\, \frac{\lambda^2}{M_H^2}\Bigl(  3 M_H^2 - 4 N_C \sum_q M_q^2 \Bigr) \label{VEL}
\end{eqnarray}
\end{widetext}

This expression looks similar to Eq. (\ref{MQQ}). However, their origins are different. For example, the condition for the cancellation of quadratic divergences relies on the identity $N_C=3$ while in the derivation of the Nambu sum rule this was never used. The Nambu sum rule Eq. (\ref{NSR}) in the models considered above works for any number of colors. Also, in Eq. (\ref{VEL}) the number of the components of the scalar is important.
Therefore, we come to the conclusion that the nature of Veltman identity Eq. (\ref{VELTMAN}) differs from the nature of the Nambu sum rule. Their coincidence at $N_C=3$ in the absence of the gauge fields is, presumably, an accident.

\section{Conclusions and discussion}
In this paper we have calculated the bosonic spectrum in the particular case of the model
suggested by Miransky and coauthors in \cite{Miransky} that involves all 6 quarks. Our model appears when the constraints on the values of the coefficients are imposed. These constraints come from the supposition that the microscopic theory approximated by the given NJL model has the large symmetry. This $U(2\times 3)_L\otimes U(2\times3)_R$ symmetry  is broken softly down to $U(2)_L\otimes U(2)_L\otimes U(2)_L\otimes U(1)_R \otimes ... \otimes U(1)_R$. In the zeroth order approximation the parameters $L_{ud},L_{cs},L_{tb},R_{ud},R_{cs}, R_{tb}$, and $I_{up}, I_{down}$ of the lagrangian Eq. (\ref{Stopcolor_}) are equal to each other, and all quarks acquire equal masses. In this approximation the symmetry  $U(2\times 3)_L\otimes U(2\times3)_R$  is preserved. In the next approximation this symmetry is violated, and the elements of matrices $L,R,I$ receive small corrections that provide the validity of the Nambu sum rule and the difference in quark masses.

 At the present moment we do not intend to consider this model as realistic. Our aim was to demonstrate how the sum rule Eq. (\ref{NSR}) emerges in relativistic models. Nevertheless, in principle, one may try to update this model in order to move it towards a realistic theory.
 In order to do this one needs to provide large masses for the light scalar bosons of this model.
 It is worth mentioning that the energy scale of the microscopic theory that has the considered NJL model as an approximation should be essentially larger than $1$ TeV. In order to pass the existing constraints on the FCNC we need $[1/g]^{1/2} \ge 10^3$ TeV \cite{Simmons}.  This implies that  $\Lambda \ge 10^3$ TeV. The large value of $\Lambda$ is also necessary in order to provide the realistic value of $F_T \approx 245$ GeV.  In addition, one must provide that the production cross sections of the composite Higgs bosons with $130\, {\rm GeV} < M_H < 550\, {\rm GeV}$ are much smaller than that of the Standard Model Higgs.

Which is even more important, - in order to produce the masses of the excitations much smaller than the cutoff, the unknown microscopic theory should provide that the trans - $\Lambda$ degrees of freedom give contributions that exactly cancel the dominant divergent higher loop contributions to the fermion and the boson masses of the effective theory given by Eq. (\ref{Stopcolor_}). Such a cancellation may occur due to the mechanism similar to that of quantum hydrodynamics \cite{quantum_hydro}. Namely, in quantum hydrodynamics  there exists the ultraviolet cutoff $\Lambda$, and the divergent contributions to vacuum energy are present. These contributions, however, are exactly cancelled by the contributions of the trans - $\Lambda$ degrees of freedom of the microscopic theory. The cancellation occurs due to the thermodynamical stability of vacuum. We imply that such a mechanism works in the unknown microscopic theory having the NJL model with action Eq. (\ref{Stopcolor_}) as a low energy approximation. This cancellation allows to use one - loop approximation to the NJL model for the calculation of various quantities just like the classical hydrodynamics can be used disregarding divergent loop contributions.

 Our analysis prompts that in the realistic model, that inherits the structure of the considered toy model, the Nambu sum rule may appear in the form of Eq. (\ref{NSR}). If so, it gives an important constraint on the bosonic spectrum.
The Nambu sum rule generalizes the relation noticed by Nambu in \cite{Nambu}. According to this sum rule the sum of the composite scalar boson masses squared (within each channel) is equal to  $2 M_f$ squared, where $M_f$ is the mass of the heaviest fermion that contributes to the formation of the given composite scalar boson. (It is implied that the single fermion dominates in the formation of this state, i.e. its mass is essentially larger than the masses of the other fermions that contribute to the given composite boson.) Originally this sum rule was considered by Nambu in $^3$He-B and in the conventional superconductivity. In the present paper we also consider how the Nambu sum rule emerges in $^3$He-A including the thin films.  We mention the analogue of this sum rule in QCD at finite chemical potential.

We feel this natural to suppose that the top quark contributes to the formation of the composite Higgs bosons.  The other composite scalar bosons would have much smaller masses. The fact that such states are not observed means that the formation of these states is suppressed. For example, the light scalar bosons may be eaten by some extra gauge fields that acquire masses due to the Higgs mechanism.  It is worth mentioning, that the Nambu sum rule alone cannot predict the masses of all composite Higgs bosons. There exist infinitely many possibilities. Below we list a few of them that seem to us interesting and instructive. In all these cases it is implied that in the r.h.s. of Eq. (\ref{NSR}) the top quark mass stands.

 \begin{enumerate}
 \item{} If there are two (doubly degenerated) Higgs bosons in the channel that contains the $125$ GeV Higgs, then the partner of the $125$ GeV boson should have mass around   $210$ GeV.

 \item{} If there are only two states in this channel, then the partner of the $125$ GeV Higgs should have the mass around $325$ GeV. Then the two Higgs masses $M_{H1} = 125$ GeV and  $M_{H1} = 325$ GeV satisfy the relations $M_{H1} = \sqrt{1/8} \, (2 M_T), \quad M_{H2} = \sqrt{7/8} \, (2 M_T)$.
These relations are to be compared with Eq. (\ref{He3B2}).

\item{} In the channel with two states of equal masses the $245$ GeV Higgs bosons should appear in analogy with $^3$He-A considered in  Section 2.  Again, a certain excess of events in this region has been observed by ATLAS in 2011 (see, for example, \cite{ATLAS}).

\item{} There is an interesting possibility that there exist 8 Higgs bosons of equal masses in a certain channel. Then the Nambu sum rule predicts
$M_H = 125$ GeV, i.e. the value of mass reported recently as the candidate for the mass of the Standard Model Higgs boson.

\item{} If in the given channel there are only two Higgs bosons, and one of them is Goldstone boson, the other one should have mass around $350$ GeV. (This is the case of the $t\bar{t}$ channel in the original model of top quark condensation by Bardeen and coauthors \cite{topcolor1}. Thus the discovery of the $125$ GeV Higgs boson excludes this model. This also excludes the majority of the technicolor models considered so far, including the so - called walking technicolor models (substitute the mass of the technifermion instead of the top quark mass into the Nambu sum rule).)
\end{enumerate}

We did not consider in this paper the possibility that the order parameter in relativistic NJL model has the structure of the space - time tensor as in $^3$He (see e.g. \cite{Akama1978,Volovik1986,Wetterich2004,Diakonov2011,Klinkhamer2012}). The simplest models of this kind appear as a modification of our toy model with the action Eq. (\ref{Stopcolor_}), where $\bar{\chi}_{\alpha}(p_+) O_{ij} \chi^{\beta}(p_-)$ stands instead of $\bar{\chi}_{\alpha,L}(p_+) \chi^{\beta}_R(p_-)$. Here $O_{ij}$ is the space - time tensor composed of gamma - matrices and momenta $p_{\pm}$ \footnote{The direct relativistic generalization of the  $^3$He model corresponds to  $O_{lm} \sim \gamma^5 \gamma^{ij} \epsilon_{ijkl} (p_-+p_+)_m (p_+-p_-)^k$.}.

\section*{Acknowledgements}
The authors kindly acknowledge useful remarks by J.D. Bjorken, D.I. Diakonov,
T.W.B. Kibble, F.R. Klinkhamer, S.Nussinov, M.I. Polikarpov,  A.M. Polyakov, M.
Shaposhnikov, V.I. Shevchenko, T. Vachaspati, M.I. Vysotsky, V.I. Zakharov. The
authors are very much obliged to V.B.Gavrilov and M.V.Danilov for the
explanation of the experimental situation with the search of new particles at
the LHC. This work was partly supported by RFBR grant 11-02-01227, by the
Federal Special-Purpose Programme 'Human Capital' of the Russian Ministry of
Science and Education, by Federal Special-Purpose Programme 07.514.12.4028. GEV
acknowledges a financial support of the Academy of Finland and its COE program,
and the EU  FP7 program ($\#$228464 Microkelvin).

\end{document}